\documentclass[useAMS]{mn2e}

\usepackage{graphicx}
\usepackage{amssymb}

\font\tenbg=cmmib10 at 10pt
\def \rvecmu{{\hbox{\tenbg\char'026}}}

\title{Bondi-Hoyle Accretion onto Magnetized Neutron Star}

\author[O.D. Toropina et al.]
{O.D. Toropina,$^1$\thanks{E-mail:toropina@iki.rssi.ru}, M.M.
Romanova,$^2$, and R.V.E. Lovelace,$^{2}$\\
$^1$ Space Research Institute, Russian Academy of Sciences,
Profsoyuznaya 84/32, Moscow 117997, Russia\\
$^2$ Department of Astronomy, Cornell University, Ithaca, NY
14853-6801}

\date{\today}
\pagerange{\pageref{firstpage}--\pageref{lastpage}}
\pubyear{2011}

\begin{document}

\label{firstpage}

\maketitle

\begin{abstract}

     Axisymmetric MHD simulations are used to investigate the Bondi-Hoyle accretion onto an isolated magnetized neutron star  moving
supersonically (with Mach number of 3) through the interstellar medium.
    The star is assumed to have a dipole magnetic field
aligned with its motion and a magnetospheric radius $R_{\rm m}$
less then the accretion radius $R_{\rm BH}$, so that the
gravitational focusing is important.
       We find that the accretion rate to a magnetized
star is  smaller than that to a non-magnetized star
for the parameters considered.
    Close to the star the accreting matter falls to the star's surface
along the magnetic poles with a larger mass flow to the
leeward pole of the star.
     In the case of a relatively large stellar magnetic field,
the star's  magnetic field is stretched in the direction of the
matter flow outside of $R_{\rm m}$ (towards the  windward
 side of the  star).
 For weaker magnetic fields we observed
oscillations of the closed magnetosphere frontward
and backward.
     These are accompanied by
strong oscillations of the mass accretion rate which varies
by factors $\sim 3$.
    Old slowly rotating neutron stars with no radio emission may
be visible in the X-ray band due to accretion of interstellar matter.
    In general, the star's velocity, magnetic moment,  and angular velocity
vectors may all be in different directions so that the accretion
luminosity will be modulated at the star's rotation rate.


\end{abstract}

\section{Introduction}

Accretion onto a supersonically moving star is an
important astrophysical process which has been studied
theoretically and with simulations over many years (review by Edgar 2004).
This type of accretion has
 applications in many areas.
    For example, the
accretion by a compact star in a binary system where the gas supply
is  from a stellar wind or from a common envelope
(Petterson 1978; Taam and Sandquist 2000); and
 flows in regions of
star formation (Bonnell et al. 2001;  Edgar and Clarke 2004).

  Supersonic accretion of a pressureless fluid onto a gravitating center was
investigated analytically by Hoyle and Lyttleton (1939) and an analytic
solution was found by Bisnovatyi-Kogan et al. (1979).   Later,
Bondi and Hoyle (1944) developed a model including the fluid
pressure.   The Bondi-Hoyle model
gives an explicit prediction for the accretion rate.

   Supersonic accretion has been studied using numerical hydrodynamic
simulations by Ruffert  (Ruffert, 1994a; Ruffert and Arnett,
1994; Ruffert, 1994b). He studied the flow of gas past an accretor
of varying sizes. For small accretors, the accretion rates
obtained were in broad agreement with theoretical predictions.
Accretion of matter to a nonmagnetized gravitating
center was investigated by Bisnovatyi-Kogan \& Pogorelov (1997).
Pogorelov, Ohsugi, \& Matsuda (2000) studied the Bondi-Hoyle
accretion flow and its dependence on parameters but also for
a nonmagnetized center.

    Here, we investigate Bondi-Hoyle accretion to magnetized
stars in application to the accretion of the
interstellar medium (ISM)  onto
Isolated Old Neutron Stars (IONS).\footnote{Isolated neutron stars have several stages in their evolution.
   Initially they  rotate rapidly and may be observed as a radio pulsars.
As the star spins down and its period increases above several
seconds it is thought that the pulsar mechnism stops working.}
     These slowly rotating neutron stars may be visible due
accretion of the interstellar medium.
These stars may have dynamically important magnetic fields.
    We use axisymmetric magnetohydrodynamics
simulations to model the accretion onto the  star.
Earlier, we
investigated cases of supersonic motion of  {\it strongly}
magnetized stars through the ISM in cases of slow rotation
(Toropina at all 2001; hereafter T01) and fast rotation in the {\it propeller}
regime (Toropina, Romanova, Lovelace 2006; hereafter T06). In
these cases the magnetospheric radius is greater then accretion
radius $R_{\rm m} \ge R_{\rm BH}$ and gravitational focusing is not
important. Both cases correspond to magnetars with
magnetic field $B \gtrsim  10^{14}$ G.

\begin{figure*}
\includegraphics[scale=.6]{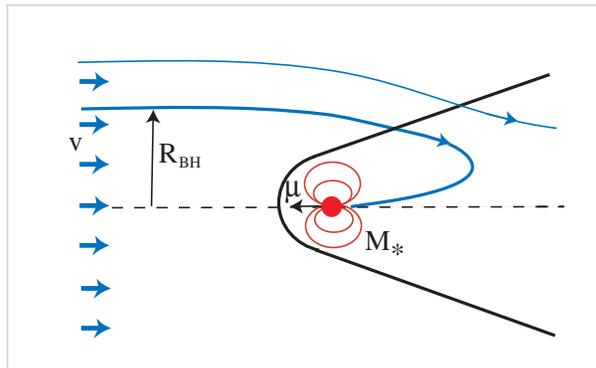}
\caption{Sketch of the Bondi-Hoyle accretion onto a supersonically
moving magnetized star.  The thick black line is the shock wave
and the red lines are the field lines of the star.} \label{Figure
1}
\end{figure*}

The present simulations are analogous to those in T01 and T06.
However, here the star has a more typical neutron star
magnetic field, $B \sim 10^{10} - 10^{12}$ G.   The
magnetospheric radius is less then the accretion radius
and gravitational focusing is important. A star gathers
matter from the region inside the accretion radius.

Sections 2 and 3  describe the physical situation and simulation
model. In Sec. 4 we discuss the results  of our  simulations. In
Sec. 5 we apply our results to more realistic neutron stars.
Conclusions are given in Sec. 6.

\section{Physical Model}

A non-magnetized star moving through the ISM  captures matter
gravitationally from the accretion or Bondi-Hoyle radius
$$
  R_{\rm BH}= {2GM \over c_s^2+v^2}\approx
3.8\times 10^{12}{ M_{1.4}  \over v_{100}^{2}}~{\rm cm}~, \eqno(1)
$$
where $v_{100}\equiv v/(100~{\rm km/s})$ is the normalized
velocity of the star, $c_s$ the sound speed of the undisturbed
ISM, and $M_{1.4} \equiv M/(1.4 M_\odot)$ is the normalized mass
of the star.
    The mass accretion rate
at high Mach numbers ${\cal M}\equiv v/c_s>>1$ was derived by
Hoyle and Lyttleton (1939),
$$
\dot M_{\rm HL}=4\pi (GM)^2 {\rho \over v^{3}}\approx 7.3\times 10^8
{n \over v_{100}^{3}} M_{1.4}^2~{\rm \frac{g}{s}}~, \eqno(2)
$$
where $\rho$ is the mass-density of the ISM and $n=n/{1~{\rm
cm}^{-3}}$ is the normalized number density.

The Hoyle-Lyttleton model neglects the fluid pressure
as mentioned.
   Later, Bondi and Hoyle (1944) extended the analysis to
include the pressure.   A more general formula for arbitrary Mach
numbers ${\cal M} = v/c_s$ was proposed by Bondi (1952),
$$
\dot M_{\rm BH} = \pi R_{\rm BH}^2 \rho v =4\pi \alpha (GM)^2 {\rho \over
(v^2+c_s^2)^{3/2}}~, \eqno(3)
$$
where  the coefficient $\alpha$ is of the order of unity.
Bondi suggested $\alpha=1/2$, but here we assume $\alpha =1$
in agreement with equation (2).
    Values of the $\alpha$ parameter for different
specific heat ratios  are determined by the simulations of
Pogorelov, Ohsugi, \& Matsuda (2000). A sketch of the
geometry for the Bondi-Hoyle analysis is shown on Figure 1.

If the fast moving star has a significant magnetic field, then the situation is more
complex.
      In the limit of a strong stellar magnetic field, the ram pressure
of ISM can be balanced by the magnetic pressure of
 the star's magnetosphere at a radius
 $R_{\rm m}$ which is larger than
 $ R_{\rm BH}$.   In this limit the ISM interacts directly with
 the star's   ``large'' magnetosphere.
      This case investigated earlier by T01 is
relevant  to stars with very strong magnetic fields
(e.g., magnetars) or to stars with very high velocity.

    In the opposite limit where $R_{\rm m} < R_{\rm BH}$
the situation is different.
     First, matter is captured by the
gravitational field of the star as in Bondi-Hoyle accretion onto a
non-magnetized star and it accumulates.
     If  the magnetosphere of the star is very small, $R_{\rm m} \ll R_{\rm BH}$,
and the gravitationally attracted matter accretes to the
star in a spherically symmetric flow, then
$$
   R_{{\rm m}}= \bigg(\frac{B_*^2 R_*^6}
{\dot M\sqrt{2GM} }\bigg)^{2/7}~, \eqno(4)
$$
(e.g., Lipunov 1992), where $B_*$ is the magnetic field at the
surface of the star of radius $R_*$ and $\dot{M}$ is the accretion
rate.

If the magnetized star accretes matter with the same rate as a
non-magnetized star, $\dot M = \dot M_{\rm BH}$, then the
magnetospheric radius is
$$
R_{\rm m}\approx 4.7\times 10^{9}\frac{B_{10}^{4/7} R_6^{12/7}
v_{100}^{6/7}} {M_{1.4}^{5/7} n^{2/7}}~ {\rm cm}~, \eqno(5)
$$
from equation (4) with $B_{10}\equiv B_*/10^{10}$G and $R_6 \equiv
R_*/10^6$cm.
    For the adopted reference parameters, $R_{\rm m}$  is roughly
equal to  $R_{\rm BH}/10^{3}$.   Our simulations discussed
below show that the accretion rate to the star is in general {\it less}
than $\dot{M}_{\rm BH}$.

 Our earlier studies of accretion onto magnetized stars
(Toropina at al. 2001; Romanova at al. 2003; Toropina at al. 2003) showed
that the magnetosphere acts as an obstacle for the flow and that the
 star accretes at a {\it lower} rate than a
non-magnetized star for the same $v, ~c_s,$ and $M$. In all those
simulations we considered strongly magnetized stars including
magnetars.

In this work we consider stars with magnetic fields $B \sim
10^{10} - 10^{12}$ G, and investigate its interaction with the
interstellar medium. We neglect rotation of the star.

\section{Numerical Model}

We  investigate the interaction of fast moving magnetized star
with the ISM using an axisymmetric, resistive MHD code. The code
incorporates the methods of local iterations and
flux-corrected-transport developed by Zhukov, Zabrodin, \&
Feodoritova (1993). This code was used in our earlier simulations
of Bondi accretion onto a magnetized star (Toropin at. al. 1999,
hereafter T99), propagation of a magnetized star through the ISM
(T01), spherical accretion to a star in the ``propeller" regime
(Romanova at al. 2003; hereafter R03) and spinning-down of moving
magnetars in the propeller regime (T06).

The flow is described by the resistive MHD equations:
$$
    {\partial \rho \over
    \partial t}+
    {\bf \nabla}{\bf \cdot}
\left(\rho~{\bf v}\right) =0{~ ,}
$$
$$
\rho\frac{\partial{\bf v}} {\partial t}
              +\rho ({\bf v}
\cdot{\bf \nabla}){\bf v}=
    -{\bf \nabla}p+
{1 \over c}{\bf J}\times {\bf B} + {\bf F}^{g}{ ~,}
$$
$$
    \frac{\partial {\bf B}}
{\partial t}
=
    {\bf \nabla}{\bf \times}
\left({\bf v}{\bf \times} {\bf B}\right)
    +
    \frac{c^2}{4\pi\sigma}
\nabla^2{\bf B} {~,}
$$
$$
    \frac{\partial (\rho\varepsilon)
}{\partial t}+
    {\bf \nabla}\cdot \left(\rho
\varepsilon{\bf v}\right)
=
    -p\nabla{\bf \cdot}
{\bf v} +\frac{{\bf J}^2} {\sigma}{~.} \eqno(6)
$$
We assume axisymmetry $(\partial/\partial \phi =0)$, but calculate
all three components of velocity and magnetic field ${\bf v}$ and
${\bf B}$.
    The equation of state is taken to be that for an ideal gas,
$p=(\gamma-1)\rho \varepsilon$, where $\gamma=5/3$ is
the specific heat ratio and $\varepsilon$ is the specific
internal energy of the gas.
      The equations incorporate Ohm's law ${\bf
J}=\sigma({\bf E}+{\bf v}  \times {\bf B}/c)$, where $\sigma$ is
the electric conductivity.
     The associated magnetic diffusivity,
$\eta_{\rm m} \equiv c^2\!/(4\pi\sigma)$, is considered to be a constant
within the computational region. In equation (2) the gravitational
force, ${\bf F}^{g} = -GM\!\rho{\bf R}/\!R^3$, is due to the
star because the self-gravity of the accreting gas is negligible.

We use a cylindrical, inertial coordinate system
$\left(r,\phi,z\right)$ with the $z-$axis parallel to the star's
dipole moment ${\rvecmu}$ and rotation axis ${\bf \Omega}$. The
vector potential $\bf A$ is calculated so that automatically
${\bf \nabla}\cdot{\bf B}=0$ at all times. We neglect rotation of
the star, ${\bf \Omega}=0$. The intrinsic magnetic field of the
star is taken to be an aligned dipole, with vector potential ${\bf
A}={\rvecmu} \times{\bf R}/{R^3}$.

We measure length in units of the Bondi radius $R_{\rm B }\equiv
{GM}/c_{s}^2={\cal M}^2R_{\rm BH}/2$, with $c_s$ the sound speed
at infinity, the density in units of the density of the
interstellar medium at infinity, $\rho$, and the magnetic field
strength in units of $B_0$ which is the field at the pole of the
numerical star.
  Here we use the term numerical star to distinguish the model star (inner boundary) from
 the actual neutron star.
 The magnetic moment is measured in units of $\mu_0
=B_0 R_{\rm B}^3/2$.

After reduction to  dimensionless form,
the MHD equations involve the dimensionless parameters:
$$
\beta \equiv \frac{8\pi p_{0}}{B_0^2}~, \quad\quad \tilde{\eta}_{\rm m}
\equiv {\eta_{\rm m} \over R_{\rm B} v_0} = {1\over Re_{\rm m}}{~,}
\eqno(7)
$$
where $\tilde{\eta}_{\rm m}$ is the dimensionless magnetic diffusivity,
$Re_{\rm m}$ is the magnetic Reynolds number, and $\beta$ is
 the ratio of the ISM pressure  at infinity $p_0$ and the magnetic pressure at the poles of the numerical star.

Simulations were done in a cylindrical region $(Z_{\rm min}\le z\le
Z_{\rm max}, 0\le r\le R_{\rm max})$. The numerical star was represented  by
a small cylindrical box with dimensions $R_*<<R_{\rm max}$ and
$|Z_*|<<Z_{\rm max}$. A uniform $(r,z)$ grid with $N_R\times N_Z$ cells was used.

Initially, the density $\rho(r,z)$ and the velocity ${\bf v}(r,z)$
are taken to be constant in the region $\rho(r,z)=\rho$ and
$v=v_{*}$, $v_\phi =0$. Also, initially the vector potential ${\bf
A}$ was taken to be that of a dipole so that $B_\phi=0$. The
vector potential was fixed inside the numerical star and at its
surface during the simulations.

The outer boundaries of the computational region were treated as
follows. Supersonic inflow with Mach number ${\cal M}$ was
specified at the upstream boundary $(z=Z_{\rm min}, 0\le r \le
R_{\rm max})$. The variables $(\rho,~v_r,~v_z,~\varepsilon)$ are fixed
with $\rho$ and $\varepsilon$ corresponding to the ISM values,
$v_z=0$, $v_r=v_{\infty}={\cal M}c_s$. The inflowing matter is
unmagnetized with ${\bf B}=0$. At the downstream boundary
$(z=Z_{\rm max}, 0\le r \le R_{\rm max})$ and at the cylindrical boundary
$(Z_{\rm min}\le z\le Z_{\rm max}, r=R_{\rm max}$, a free boundary condition
was applied, $\partial/\partial{\bf n}=0$. The boundary conditions are
described in greater detail in T01 and T06.

The size of the computational region for most of the simulations
was $R_{\rm max}=0.27$, $Z_{\rm min}=-R_{\rm max}=-0.27$, with
$Z_{\rm max}=2R_{\rm max}=0.54$ in units of the Bondi radius. The
theoretically derived accretion radius (Bondi-Hoyle radius)
$R_{\rm BH}=0.2$. Therefore the size of the computational region
is sufficient for our simulations.

The grid $N_R\times N_Z$ was $1297\times 433$ in most of cases.
The radius of the numerical star was $R_*=0.0025$ in all cases.

\section{Results}

We consider Bondi-Hoyle accretion onto a star
with a small magnetosphere.
    Simulations were done for
a number of values of the star's  magnetic moments $\mu$.
     The numerical star (inner boundary) is typically much larger
than the true radius of the neutron star.
      Thus the actual star is
``hidden'' inside the numerical star.
    The Mach number for all of the present simulations
is ${\cal M}=3$.
    Also,  $\beta=10^{-6}$,
which corresponds to the magnetic moment $\mu=10^{30}$Gcm$^3$.   The magnetic diffusivity is
$\tilde{\eta}_{\rm m}=10^{-5}$.
    We measure time in  units of
$t_0=(Z_{max} - Z_{min})/v_\infty=0.81R_{\rm B}/v_\infty$, which is the crossing time of
the computational region with the star's velocity, $v_{\infty}$.

\begin{figure*}
\includegraphics[scale=.7]{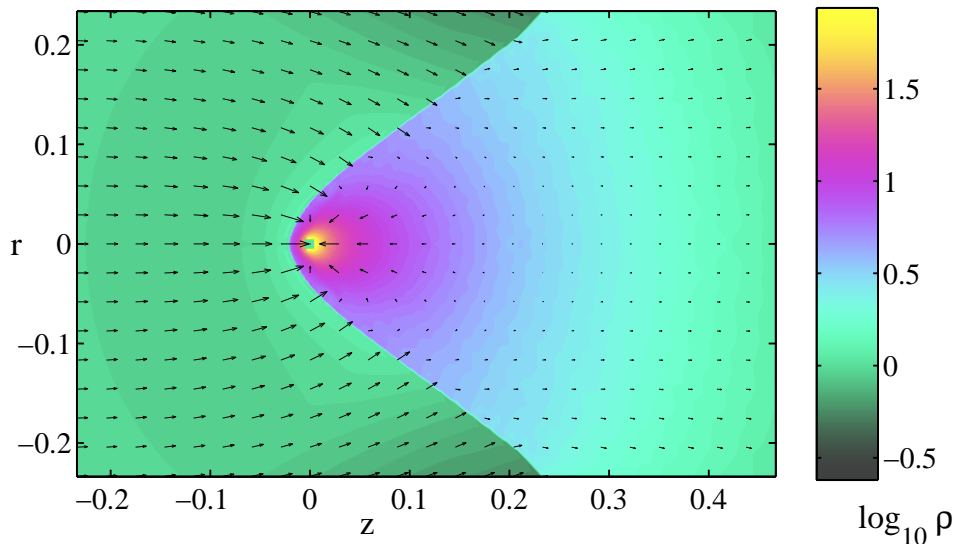}
\caption{A hydrodynamic simulations of the BHL accretion to a {\it
non-magnetized} star for Mach number ${\cal M}=3$.
   The central region of the flow is shown at a late time ($t= 7.0 t_0$)
   when the flow is stationary. The background colors   represent the logarithm of
the density.   The arrows represent velocity vectors. Distances
are measured in units of the Bondi radius $R_{\rm B}=GM/c_s^2$.}
\label{Figure 2}
\end{figure*}

\begin{figure*}
\includegraphics[scale=.7]{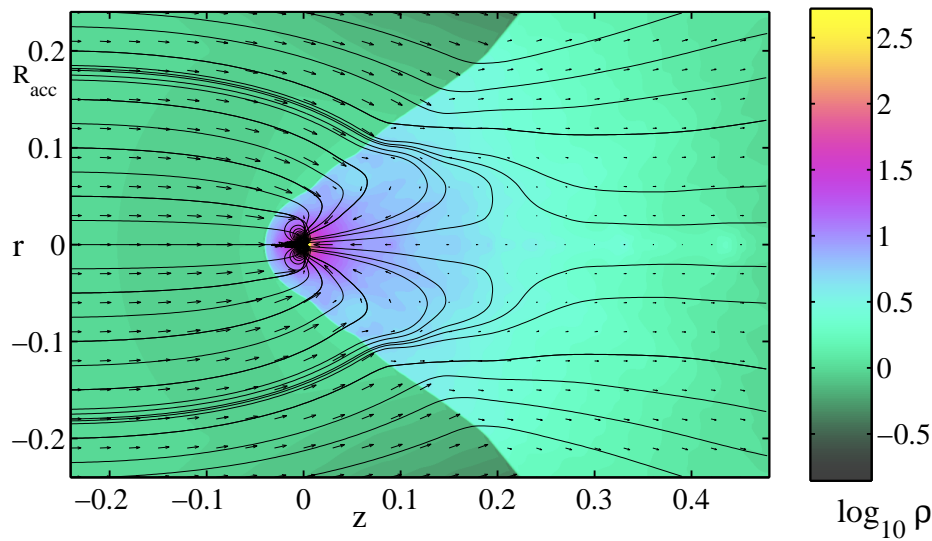}
\caption{Matter flow around a weakly magnetized star moving
through the interstellar medium with Mach number ${\cal M}=3$ at a
late time $t= 4.4 t_0$. The background represents the logarithm of
density and the solid lines are streamlines. The length of the
arrows is proportional to the poloidal velocity} \label{Figure 3}
\end{figure*}

\subsection{Classical Bondi-Hoyle Accretion}

To test our model we performed hydrodynamic simulations of the
Bondi-Hoyle accretion to a {\it non-magnetized} star for Mach
number ${\cal M}=3$.  Dimensionless accretion radius for this
parameter $R_{\rm BH}=0.2$.    The star collects matter from
inside this radius.   We verified that  the flow is close to that
described earlier in T01 where matter forms a conical shock wave.

Figure 2 shows the main features of the flow at a late time $t=
7.0 t_0$ when the flow is stationary.
    We calculated the accretion rate $\dot M$ to the numerical star
and got a value $\dot{M} \approx 0.7 \dot{M}_{\rm BH}$ for numerical
star $R_*=0.0025$ were $\dot{M}_{\rm BH}$ is   the Bondi-Hoyle accretion rate of equation (3) with $\alpha=1$.
  In our earlier calculation we got a slightly
smaller value, $\dot M\approx 0.5
\dot{M}_{\rm BH}$ for a numerical star $R_*=0.02$ (T01). This
difference can be explained by the higher resolution of the
present calculations. And this difference agrees with Ruffert's
results on the dependence of $\dot M$ on  numerical star size for
the sizes used (Ruffert 1994).

\begin{figure*}
\includegraphics[scale=0.7]{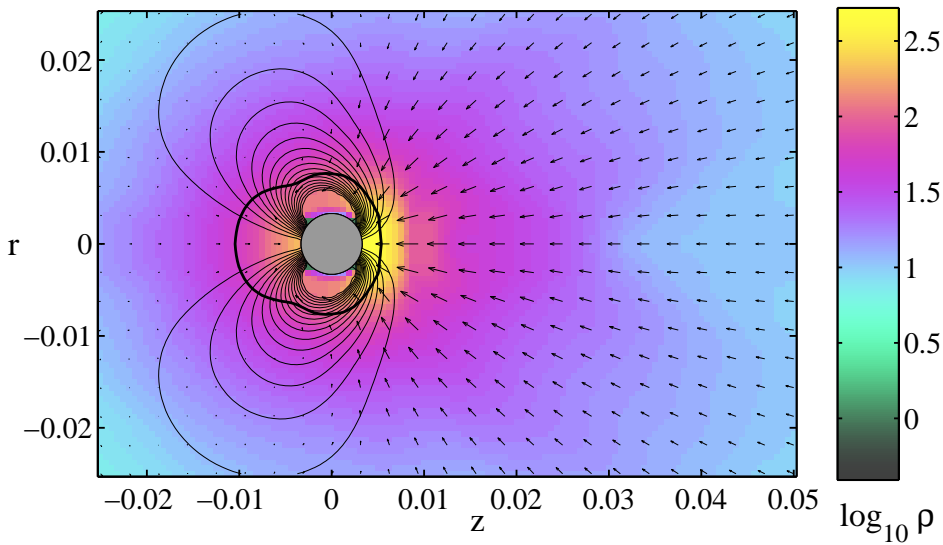}
\caption{A central region around a weakly magnetized star moving
through the interstellar medium with Mach number ${\cal M}=3$ at a
late time $t= 4.4 t_0$.  Zoomed-in view of the stellar magnetic
field which is deformed by the flow. The thick black line
surrounding the star is the magnetospheric radius $r_m$. The gray
circle is the numerical star.} \label{Figure 4}
\end{figure*}


\subsection{Bondi-Hoyle accretion to a star with a small magnetosphere}

We performed simulations for a number of values of the magnetic
moment.
  Initially we chose $\mu=10^{30}$Gcm$^3$ (dimensionless moment $\mu=1.6 \times 10^{-8}$), which
corresponds to a star's surface magnetic field $B=10^{12}$G.
We investigated this case in detail.

     Figure 3 shows the main features of the flow.
 Because of the magnetospheric radius $R_{\rm m} << R_{\rm BH}$, the
gravitational focusing is important and the star gathers matter
from the region inside the accretion radius $R_{\rm BH} \approx 0.18$,
 which corresponds approximately to the
theoretically derived accretion radius (Bondi-Hoyle radius)
$R_{\rm BH}=0.2$.
    The magnetospheric radius $R_{\rm m}=0.02$ is about $10$ times smaller than the gravitational radius $R_{\rm BH}$.
     At the same time the average magnetospheric radius
$R_{\rm m}$  is about $3$ times larger than the numerical star radius
$R_*=0.0025$.
   The star's magnetic field at  distances $r>R_{\rm m}$ is significantly
affected by the matter flow.
   An excess of matter flows from the backside
of the star (return flow) and consequently
the  magnetic field lines are pushed towards the
``front'' of the star forming ``ear''-like shapes.
      The ``ears" oscillate in the $z-$direction. The conical
shock wave is close to that found in the
hydrodynamic case which is expected
 since the Mach numbers are the same.

     Figure 4 shows zoomed-in view of the stellar magnetic which
is deformed by the flow at a late time $t=4.4t_0$.   The thick
black line surrounding the star is the magnetospheric
radius $r_m$ where $\rho {\bf v}^2  = {\bf B}^2/8\pi$.

Figure 5 shows the dependences of the density, velocity and temperature
along the $z-$axis. There is a normal shock wave at
$z\approx -0.04$.    The density
increases as the star is approached; and it is zero in
the region of the star where matter is accreted.
Behind the star ($z>0$) there is a strong density
enhancement owing to the gravitational focusing.
     At $t= 4.4 t_0$ the
density of matter behind the star is significantly
larger  than the density on the
front side.

The velocity $v_z$ decreases sharply in the shock wave to
subsonic values, but later increases again in the polar column.
Behind the star the velocity is negative in the region,
where accretion occurs and also increases again in the polar
column to  supersonic values.
   The plasma temperature $T$ and pressure $p$
increase strongly  approaching  the star along the
$\pm z$ directions.   This is due to the adiabatic
compression of the matter.

Figure 6 shows the variation of the plasma pressure $p$,  magnetic
pressure $p_{\rm mag}$ and $\beta \equiv p/ p_{\rm mag}$  along the $z-$axis.
The dashed lines correspond to the initial distributions.
   One can
see that the pressure $p$ gradually increases along $z-$axis,
has a maximum in polar columns, and then gradually decreases.
The magnetic pressure $p_{\rm mag}$ initially has a dipole shape, but
it is significantly modified by the inflow of matter.
Outside of the magnetosphere the matter pressure dominates.

\subsection{Oscillations}

We performed a number of simulation runs with different dimensionless
magnetic moments ranging from  $\mu=4.7 \times 10^{-9}$ to
$\mu=1.2 \times 10^{-7}$.
    For relatively small magnetic moments, $\mu< 2\times 10^{-8}$,
we observed oscillations of the magnetic field configuration as
shown in Figure 7 for our main case where $\mu=1.6 \times
10^{-8}$.
    These oscillations are connected with the balance
between back flow of the accreting matter
and  the tension in the distorted stellar magnetic field.
Figure 7 illustrates time evolution of magnetic field
lines.
     The field oscillations are accompanied by strong oscillations
in the accretion rate to the star as shown in the top
panel of Figure 8 for our main case where $\mu=1.6\times 10^{-8}$.
  For larger values of $\mu$ the oscillations are much
  weaker as shown in the bottom panel of Figure 8.

       A Fourier
analysis of the accretion rate oscillations for our main case gives the
 quasi-period of these oscillations
 $\Delta T \approx 0.2 t_0=0.16R_{\rm B}/v_\infty=0.72R_{\rm BH}/v_\infty$.

       Note that  in  hydrodynamic simulations Ruffert (1994b, 1996)
observed a ``flip-flop'' instability of the accretion flow in 2D
simulations and mass accretion rate fluctuation for smaller accretors in 3D
simulations.   In our simulations we see instability
of the flow in case of weak or moderate magnetic field. This
instability disappears as the magnetic field increases.
    Further,  this
instability is absent  in our purely hydrodynamic simulation with the
same  parameters.

\begin{figure*}
\includegraphics[scale=0.6]{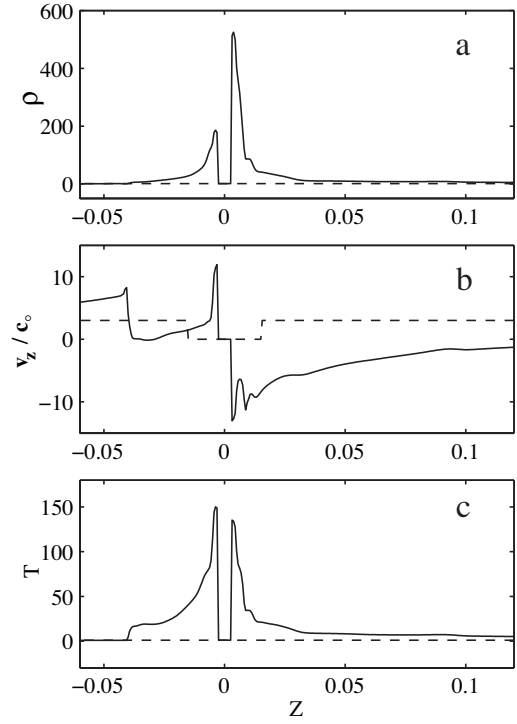}
\caption{Panel (a) shows the density, panel (b) the axial velocity
and panel (c) the temperature variation along the $z-$axis for our
main case at $t=4.4t_0$. The dashed lines give the initial
distributions.} \label{Figure 5}
\end{figure*}

\begin{figure*}
\includegraphics[scale=0.6]{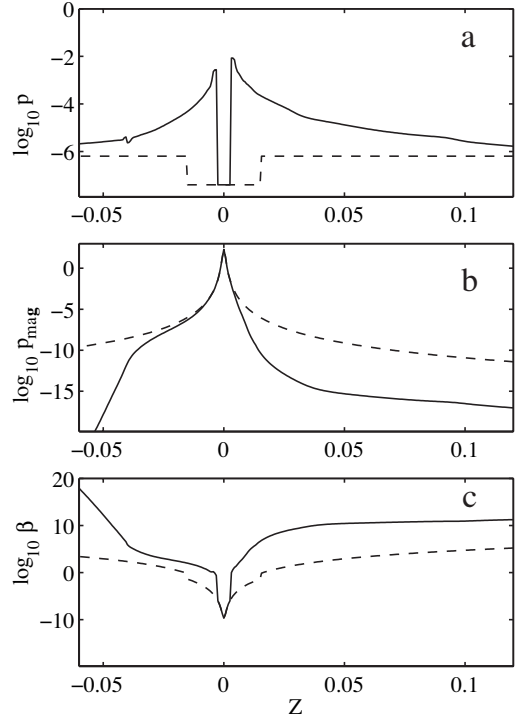}
\caption{Panel (a) shows the matter pressure $p$,  panel (b) the
magnetic pressure $p_{\rm mag}$, and panel (c) $\beta =p /p_{\rm
mag}$ variation along the $z-$axis for our main case at
$t=4.4T_0$. The dashed lines give the initial distributions.}
\label{Figure 6}
\end{figure*}

\subsection{Accretion rate}

     For our main case the mass accretion rate
is approximately
$\dot{M} \approx 0.45 \dot{M}_{BH}$.
This  value is smaller than mass accretion rate in the hydrodynamic
case with the same  parameters $\dot{M} \approx 0.7
\dot{M}_{BH}$.
      Hence even a small stellar magnetic field acts to
reduce  the mass accretion rate.
    This is in agreement with the conclusions of our
earlier work (T99, T03).

Figure 9 shows the observed
dependence of the accretion
rate on the star's magnetic moment.
    It is approximately  $\dot M \sim
\mu^{-0.4}$.
     Note however that as $\mu$ increases, the magnetospheric
radius $R_{\rm m}$ can approach the accretion radius $R_{\rm BH}$ and this can suppress the accretion.

\begin{figure*}
\includegraphics[scale=1.0]{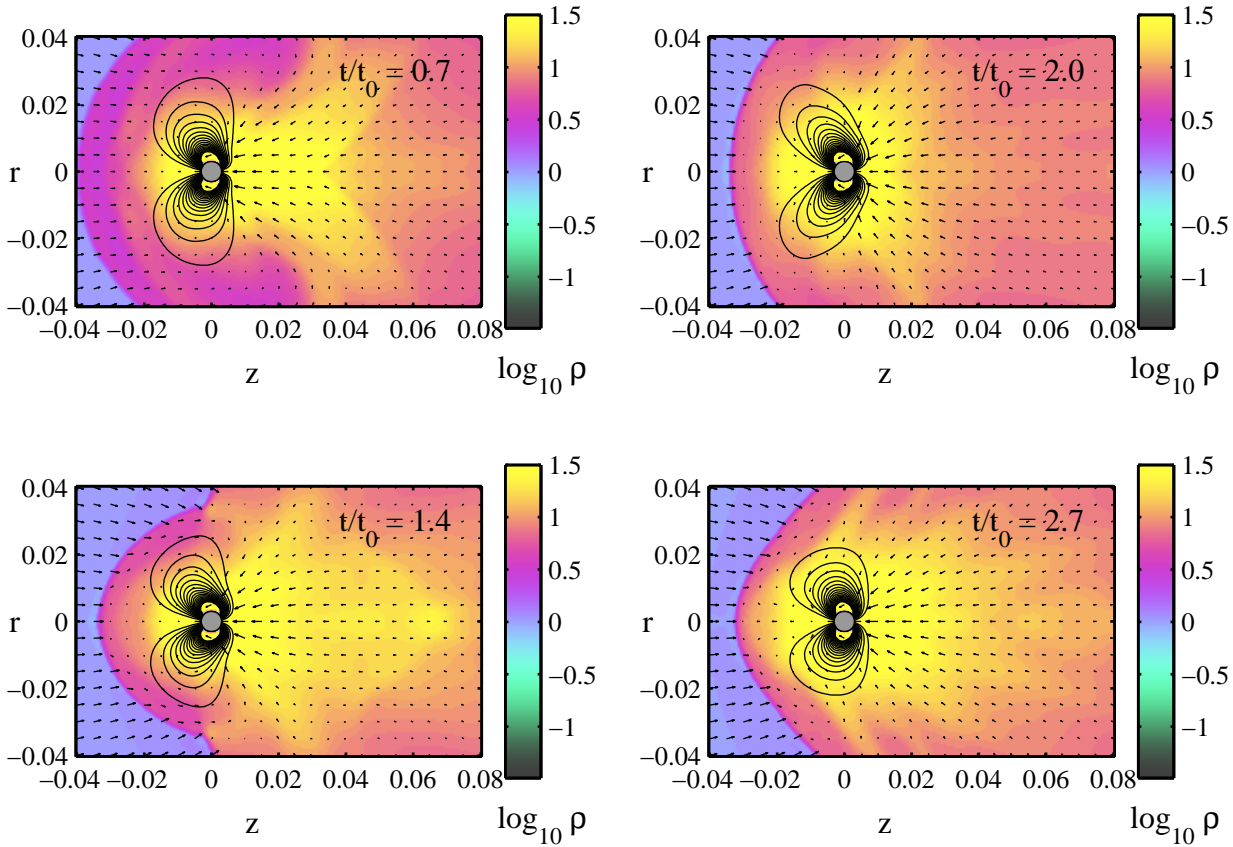}
\caption{Matter flow around a weakly magnetized star moving with
Mach number ${\cal M}=3$ at the times indicated on the plots.  The
background color represents the logarithm of density and the
length of the arrows is proportional to the poloidal velocity.The
solid lines are magnetic field lines. Distances are measured in
units of the Bondi radius.} \label{Figure 7}
\end{figure*}

\section{Astrophysical parameters}

Let us consider an isolated old neutron star (IONS) moving through
the interstellar medium.
      The star is magnetized and has a mass
$M=1.4~M_{\odot}$.
    The density of the interstellar matter is assumed to be
$\rho_\infty=1.7 \times 10^{-24}$ ~g$/$cm$^3$
($n_\infty=1/$~cm$^{3}$).
    The sound speed in the ISM is assumed to be
 $c_s=12$km/s  so that at a Mach number ${\cal M}=3$ the star moves
 with a velocity $v=36$ km/s which corresponds to a relatively slow
 moving neutron star (Arzoumanian, Chernoff, \& Cordes 2002).

The Bondi-Hoyle radius is $ R_{\rm BH} =GM_*/v^2 \approx 1.3
\times 10^{13} {\rm cm}.$
 The Bondi-Hoyle accretion rate is the $\dot{M}_{\rm BH} \approx
 5\times   10^{-17} M_\odot$/yr for a non-magnetized star.
 This gives an accretion energy release rate of $L_{\rm acc}=GM\dot{M}/R_*\approx 6\times 10^{29}$erg/s for $R_*=10^6$cm and a non-rotating star.  Both $\dot{M}$ and
 $L_{\rm acc}$ are reduced for magnetized star as as shown in Figure 10.
With $\beta=8\pi p_{\infty}/B_0^2$ and $p_\infty=\rho_{ISM}
c_s^2/\gamma$, we obtain the magnetic field at the surface of the
numerical  star $B_0 = (8 \pi /{\gamma}{\beta})^{1/2} \rho
^{1/2} c_s  \approx 0.006$G.

Recall that the size of the numerical star is $R_*= 0.0025
R_{\rm B}\approx 3\times 10^{11}{\rm cm}$ in all
simulations. Thus the magnetic moment of the numerical star is
$\mu =B_0 r_*^3/2\approx  10^{30}$
Gcm$^3$. Note that this is of the order of the magnetic moment of
typical spin powered radio pulsars which have a surface field $B_{\rm NS} \approx 10^{12}$ G and radius  of the order of $10^6$ cm.
    Thus  we can suggest that an actual neutron
star exists  inside of our numerical star.

We can estimate the radius of the magnetosphere using equation (5) as
$R_{\rm m} \approx 6.2 \times 10^{11}{\rm cm}$.
  Thus $R_{\rm m} << R_{\rm BH}$.

\begin{figure*}
\includegraphics[scale=0.7]{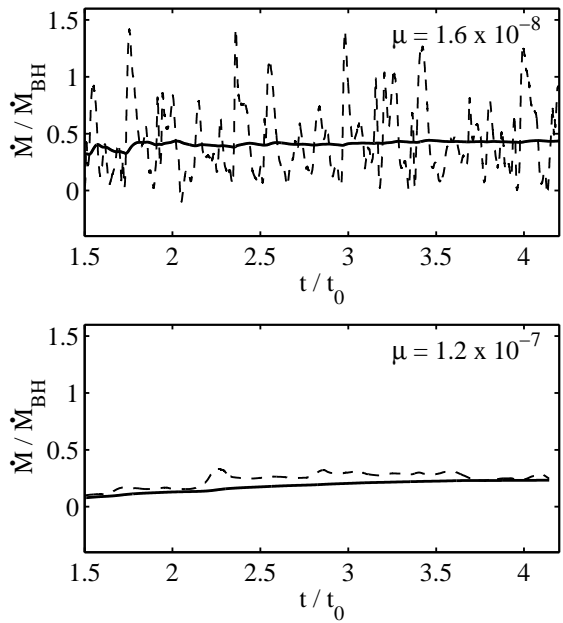}
\caption{The top panel shows the Dependence of mass accretion rate
$\dot M$ on time for our main case with $\mu=1.6\times 10^{-8}$
and the bottom panel a case with a stronger magnetic field,
$\mu=1.2\times 10^{-7}$
   The dashed lines give the mass accretion rate
normalized in Bondi-Hoyle rate,  while the solid lines give the
integrated mass flux. Time is measured in the crossing time units,
$\Delta z/v_\infty$.} \label{Figure 8}
\end{figure*}

\begin{figure*}
\includegraphics[scale=.5]{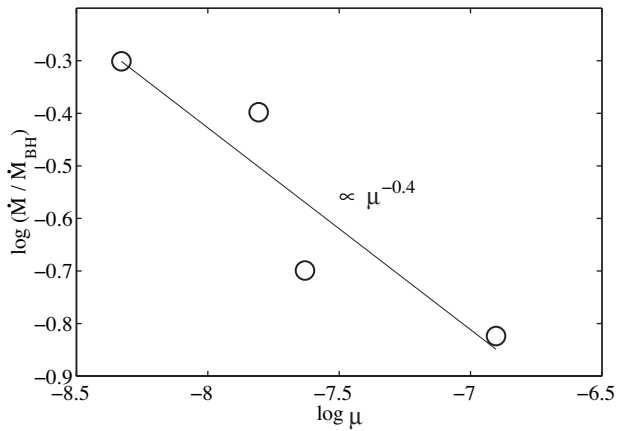}
\caption{Dependence of the mass accretion rate $\dot M$ on
magnetic moment $\mu$.  The straight line is a least-squares
straight line fit to the points.} \label{Figure 9}
\end{figure*}

\section{Conclusions}

We carried out axisymmetric MHD simulations of Bondi-Hoyle
accretion onto a fast moving neutron star (${\cal M}=3$) with an
aligned dipole magnetic field for cases where the Bondi-Hoyle
accretion radius $R_{\rm BH}$  is much larger than the star's
magnetospheric radius $R_{\rm m}$.    From this we conclude that:

1.~ Interstellar matter is captured by the star if it has
an impact parameter less than
the accretion radius $R_{\rm BH}$.
    It then falls onto the star mainly
onto the magnetic poles of the star.  More matter falls downstream
of the star.
    The accretion rate to the star is  smaller
than that to a non-magnetized star.

2.~ In case of a strong stellar magnetic field ($R_{\rm m} >
R_{\rm BH}$) the field is stretched in the direction of the
dominant flow which towards the front side of the star.

    In the case of  a weak magnetic field ($R_{\rm m} < R_{\rm
BH}$), the star's field oscillates in the front/back direction.
These oscillations are accompanied by strong
 oscillations of the  mass accretion rate. The time
scales of the oscillations is $\sim 0.7R_{\rm BH}/v_\infty$.

3.~ The accretion rate  decreases with increasing magnetic
moment approximately as $\dot M \sim \mu^{-0.4}$.

4.~    Old slowly rotating neutron stars with no radio emission may
be visible in the X-ray band due to accretion of interstellar matter.
For a neutron star moving through the interstellar medium
with  Mach number of $3$ we estimate the accretion luminosity
as $\lesssim 10^{30}$erg s$^{-1}$.
    In general, the star's velocity, magnetic moment,  and angular velocity
vectors may all be in different directions so that the accretion
luminosity will be modulated at the star's rotation rate.

\section*{Acknowledgements}

We thank  Dr. V.V. Savelyev for the development of the
original version of the MHD code used in this work and Dr. Yuriy
Toropin for discussions.
This work was supported in part by NASA grant NNX10AF63G,
NSF grant AST-1008636, RFBR
grant 11-02-00602, and Leading scientific school  grant NS-3458.2010.2
and by Russian Academy of Science program ``Origin, Structure and
Evolution of the Universe Objects''.

\clearpage

\end{document}